\documentclass{aa}
\usepackage[varg]{txfonts}
\usepackage{placeins}
\usepackage{natbib}
\usepackage{lscape}
\bibpunct{(}{)}{;}{a}{}{,} 
\usepackage{makecell}

\usepackage{changes}
\usepackage{graphicx}
\usepackage{txfonts}
\usepackage{hyperref}
\usepackage{xcolor}

\begin{document}

   \title{Impact of satellite trails on H.E.S.S. astronomical observations\thanks{Table A.2 is only available in electronic form
at the CDS via anonymous ftp to \href{cdsarc.cds.unistra.fr}{cdsarc.cds.unistra.fr (130.79.128.5)}
or via \href{https://cdsarc.cds.unistra.fr/cgi-bin/qcat?J/A+A/}{https://cdsarc.cds.unistra.fr/cgi-bin/qcat?J/A+A/}}}

   \author{T. Lang
          \inst{1}
          \and
          S.T. Spencer
          \inst{1,2}
          \and
          A.M.W. Mitchell \inst{1}
          }

   \institute{Friedrich-Alexander-Universit{\"a}t Erlangen-N{\"u}rnberg, Erlangen Centre for Astroparticle Physics, Nikolaus-Fiebiger-Str. 2, D 91058 Erlangen, Germany
         \and
             Department of Physics, Clarendon Laboratory, Parks Road, Oxford, OX1 3PU, United Kingdom
             }

 
  \abstract
   {
    The number of satellites launched into Earth's orbit has almost tripled in the last three years (to over 4000) due to the increasing commercialisation of space. Multiple satellite constellations, consisting of over 400,000 individual satellites, have either been partially launched or are proposed for launch in the near future. Many of these satellites are highly reflective, resulting in a high optical brightness that affects ground-based astronomical observations. Despite this caveat, the potential effect of these satellites on gamma-ray-observing Imaging Atmospheric Cherenkov Telescopes (IACTs) has largely been assumed to be negligible due to their nanosecond-scale integration times. However, this assumption has not been verified to date.
}
   {As IACTs are sensitive to optical wavelength light, we aim to identify satellite trails in data taken by the High Energy Stereoscopic System (H.E.S.S.) IACT array. In particular, this study is aimed at  quantifying the potential effects on data quality and extensive air shower event classification and reconstruction.}
   {Using night sky background measurements from H.E.S.S., we determined which observation times and pointing directions are affected most by these satellite trails. We then evaluated their impact on the standard Hillas parameter variables used for event analysis.}
   {Due to the brightest trails, false trigger events can occur, however, for most modern analyses, the effect on astronomical results will be minimal. We observe a mild increase in the rate of trail detections over time (approximately doubling in three years), which is partially correlated with the number of satellite launches. Overall, the fraction of H.E.S.S. data affected ($\sim0.2\%$ of dark time observations) is currently minimal. We note that these trails could still have a non-negligible effect on future Cherenkov Telescope Array observations if advanced analysis techniques designed to lower the energy threshold of the instrument are applied.}
   {}

   \titlerunning{Satellite trails in H.E.S.S. data}
   \authorrunning{T. Lang et al.}
   \maketitle
%

\section{Introduction}
Satellite constellations are designed to provide internet access across the globe \citep{mcdowell}, with several companies planning to launch $\sim10^3-10^4$ satellites into low Earth orbit (LEO) each. Two of the biggest operational satellite constellations are Starlink, which aims to launch up to 42000 satellites (with over 4200 already launched), and OneWeb, which plans to launch over 7000 satellites (620 already launched). These megaconstellations, along with others such as E-Space (337323 planned) and Guowang (12992 planned), will significantly increase the number of satellites in orbit in the near future and could significantly impact ground-based astronomical observations \citep{mcdowell}.

The effect of these megaconstellations on such observations has been widely studied in the radio and optical bands. In particular, the Zwicky Transient Facility (ZTF) recently measured the effect of visor-based modifications on the optical brightness of Starlink satellites (proposed by SpaceX to reduce their impact on astronomical observations) and found that the historically used unvisored satellites at 550km exhibit magnitudes ranging from 4.58-5.16 in the i through g bands, decreasing to 5.91-6.77 when visors are deployed.  \citep{Mr_z_2022}. The effect on future Square Kilometer Array (SKA) radio observations has also been quantified by \citet{SKA_2020_Cass}; these authors found a decrease in signal strength in the 5b band ($8.3-15.4\,\mathrm{GHz}$) of 70\% due to the satellite downlink transmission. This could render the entire band unusable if a sufficient number of satellites ($\sim100,000$) are launched.

To date, the effect of satellite trails upon ground-based gamma-ray telescopes has not been studied with experimental data. And so, in this paper, we quantify the effect of satellite trails on Imaging Atmospheric Cherenkov Telescopes (IACTs) for the first time using data from the High Energy Stereoscopic System (H.E.S.S.) array. These telescopes are designed to observe the gamma-ray sky above $\mathrm{20\,GeV}$ by observing the (optical to near-UV) Cherenkov light produced when a gamma-ray initiates an Extensive Air Shower (EAS) in the atmosphere. They observe this Cherenkov light by focusing it using large mirrors onto very-high-speed cameras that have focal planes instrumented with photomultipliers (as the light arrives on the ground within a roughly $\mathrm{10\,ns}$ window). When sufficient Cherenkov light in combinations of these photomultiplier pixels is detected, the cameras trigger and read out the photomultiplier waveform (which typically have  durations in the 10s of ns). Subsequent calibration steps and integration of the waveforms enables a pixel intensity quantified in photoelectrons (p.e.) to be obtained. 

These images are then cleaned, removing some (but not all) of the background photons from the night sky, including those from satellites. Typically a two-threshold scheme (known as `tailcut' cleaning) is employed. This includes pixels in the analysis if they exceed an intensity threshold, or if they are adjacent to such a pixel and have an intensity above a second, lower threshold. So-called Hillas parameters \citep{1985ICRC....3..445H} are then used to characterise the remaining intensity distribution. Three tasks must then be performed to determine the properties of the incident $\gamma$-ray: rejection of a large EAS background from hadronic cosmic rays, reconstruction of the direction of the primary particle, and estimation of its energy. This is performed using sophisticated Monte Carlo simulations, which can then be used to generate lookup tables and templates to infer event characteristics from observed EAS. As hadronic cosmic rays experience strong nuclear interactions in the atmosphere, they can be rejected due to producing wider and less elliptical images. Similarly, the total number of p.e. observed (Hillas amplitude) is proportional to the primary particle energy and the impact distance from the telescope, and reconstructors trained using simulated gamma-like events can therefore be used to infer the primary particle energy. IACTs typically observe EAS stereoscopically; this allows for improved background rejection against secondary muons from hadronic EAS (by requiring more than one camera to trigger for an event to be recorded) and for the incident particle direction to be reconstructed reliably by intersecting the primary axes of the Hillas ellipses \citep{Benbow:2005wj}. Lookup tables can be used to stereoscopically combine the Hillas parameters relevant for event classification (length and width). Sophisticated machine-learning and template-based techniques are now commonly used for this event classification and reconstruction \citep{Ohm_2009,Parsons2014,parsons2015hess}, but share the same underlying principle. For further information on the IACT technique, we refer to \citet{2004HESScameras, Benbow:2005wj,2006HESSCrab,Parsons2014}.

The H.E.S.S. array is made up of five IACTs located in the Khomas Highland of Namibia. It consists of four 12-meter-diameter telescopes (CT1-4, constructed 2004) and a fifth telescope (CT5, constructed 2012) with a 28-meter diameter. The addition of CT5 extends the observable energy range down to 20\,GeV and increases the sensitivity overlap with space-based instruments, such as the \textit{Fermi} Large Area Telescope \citep{Fermi}. The original CT1-4 (HESS1) cameras were upgraded between 2015-17 \citep{hess1uupgrade}, during which all components apart from the photomultipliers were replaced (henceforth referred to as HESS1U). The CT5 telescope was upgraded in 2019 with a replacement camera based on the new-generation FlashCam Cherenkov Telescope Array (CTA) design \citep{Flashcamicrc2015}. The aim of this upgrade was to further reduce the energy threshold of CT5 observations, and improve instrumental stability \citep{FlashcamPerformance}. The basic properties of the current H.E.S.S. instrumental configuration are listed in Table \ref{table:hess}. CTA will consist of at least $64$ telescopes from three IACT classes (large, medium, and small) on two sites: one on the Spanish island of La Palma, and another near Cerro Paranal in Chile. Cameras based on the FlashCam design will be deployed on the Medium Sized Telescopes on the southern site. It is therefore important to quantify the influence of satellite trails in current generation instruments in order to predict their potential influence on CTA observations in the 2030s.
\begin{table*}
\centering
\caption{Properties of the current (post-2019) H.E.S.S. instruments.}
\label{table:hess}
\begin{tabular}{c c c}     
\hline\hline
Instrument Property & CT1-4 Value & CT5 Value\\ \hline
Optical Design & Davies-Cotton & Parabolic\\
Focal Length & $\mathrm{15\,m}$ & $\mathrm{36\,m}$\\
Mirror Diameter & $\mathrm{12\,m}$ (Hexagonal) & $\mathrm{28\,m}$ (Circular Equivalent)\\ 
Number of Pixels & $\mathrm{960}$ & $\mathrm{1758}$\\
Pixel Size (On Camera) & $\mathrm{42\,mm}$ & $\mathrm{50\,mm}$ \\
Pixel Size (On Sky) & $\mathrm{0.16\,^{\circ}}$ & $\mathrm{0.067\,^{\circ}}$ \\
Average Field of View & $\mathrm{5.0\,^{\circ}}$ &  $\mathrm{3.4\,^{\circ}}$ \\
Optical PSF & $\mathrm{\sim0.022-0.035\,^{\circ}}$ & $\mathrm{\sim0.022\,^{\circ}}$\\\hline
\end{tabular}
\tablefoot{The optical point spread function (PSF) for each telescope varies, depending on the time since the last mirror alignment.}
\end{table*}

The remainder of this paper is structured as follows. In Section \ref{sec:ident},  we present our novel technique for identifying the presence of satellite trails in IACT observations using Night Sky Background (NSB) data produced by the CT5 camera. In Section \ref{sec:det}, we analyse the statistical properties of these satellite detections and attempt to quantify their effect on H.E.S.S. gamma-ray observations in Section \ref{sec:impact}. In Section \ref{sec:pred}, we predict the possible impact of such satellite trails on upcoming instruments. In Section \ref{sec:conc}, we weigh the adverse effects on our data from satellite constellations against their potential societal benefits and draw our conclusions.

\section{Data selection and methods}
\label{sec:ident}

In the IACT analysis, satellite trails are a component in NSB, the light detected by IACTs not attributed to Cherenkov emission. The largest components in NSB are starlight, scattered moonlight and airglow emission from the atmosphere; for a recent study, see \citet{stsnsb}. In principle, there are three categories of NSB phenomena that can cause transient, trail-like features in IACT data: satellites, meteors, and aircraft such as planes. First considering these other sources of trails; as meteors traverse the camera field of view (FoV) near-instantaneously, they can be removed from the analysis by performing a simple trail duration cut (further discussed in Section \ref{sec:det}). Additionally, for this analysis, we consider the likelihood of planes overflying the H.E.S.S. site in the Khomas Highland, at night, within the telescope FoV to be minimal (although this is difficult to verify as light aircraft near the H.E.S.S. site cannot be tracked). We aim to perform a statistical analysis of the trails only. Identifying the individual satellites responsible for trails over several years is both technically challenging, as two line-element files detailing satellite orbits are only valid for $\sim$2 weeks \citep{Riesing2015OrbitDF} and, irrelevant for the purposes of IACT data quality monitoring. The optical brightness of individual satellites also varies over time, due, for instance, to changes in the satellite altitude, configuration, and Starlink visor deployment \citep{2022ApJ...924L..30Mroz}. Also, this parameter has not been measured for the vast majority of objects, so we consider the identification of individual satellites to be beyond the scope of this paper.

H.E.S.S. and other current generation IACT arrays typically observe in runs, observing a single source with parallel telescope pointing for a duration of 20-30 minutes. This allows for sufficient event statistics to be collected to calculate the background event rate, whilst preventing the observational conditions from varying too much. We consider 4715 runs of H.E.S.S. data taken between the 28th of October 2019 and the 2nd of February 2023. We select only runs where all five H.E.S.S. telescopes are operational for ease of data processing, and additionally employ the standard observation run selection criteria used in in the H.E.S.S. \textit{Heidelberg Analysis Pipeline (HAP)} reconstruction chain. Runs must normally pass these criteria to be used in astrophysical analyses. This excludes runs that are 1) taken under poor atmospheric conditions, 2)  where the system trigger rate is insufficiently stable (e.g. during cloudy weather), 3)  too short in duration, 4)  poor in terms of the tracking performance, and 5)  where the cameras have large numbers of disabled pixels. A common reason for a pixel being disabled is that the photomultiplier high voltage has been turned off in response to a high flux to protect the instrument (this begins to happen at approximately $m_V=5$ for H.E.S.S. cameras, but it is a function of observational conditions). As such, this disabled pixel run selection cut will have (slightly) reduced the overall number of satellites we detected, since bright satellites will cause pixels to turn off. However, this is a small effect as runs with many disabled pixels are typically affected by hardware failures. 

We also excluded observations pointed within a five degree radius of the particularly bright NSB regions Eta Carinae and the Large Magellanic Cloud, as this sky brightness affects the performance of our track isolation algorithm. Eta Carinae's high UV brightness is also known to cause false trigger events in H.E.S.S. data \citep{etacar}. We also excluded data from February 2020 from our analysis, which refer to the number of trails detected over time, as the vast majority of runs during this observing period are of Eta Carinae and so, including this month was found to bias our results. Partial moonlight observation runs, whilst likely to play a major role in CTA observations \citep{scienceCTA}, although they have been performed by H.E.S.S. since 2020 \citep{tomankova_moonlight}, are not included in our analysis due to their higher associated NSB rates. We use the \textit{Hillas0510} tailcut cleaning scheme, where the thresholds are 10\,p.e. and 5\,p.e. respectively, for this analysis as values obtained using it are used for high-level-parameter cuts later in the \textit{HAP} reconstruction chain.

Satellite identifications in conventional IACT images have traditionally been limited to maps of disabled pixels over the camera plane. However, such disabled pixels can easily be caused by bright stars as well, and only occur in the cases of the brightest satellites. The time constant for over-illumination detection in the current CT5 camera is also configured to prevent pixels turning on and off within short time intervals; hence, satellite trails cannot be seen in deactivated pixel data for this camera. We therefore developed a new approach that appears to be able to detect trails in CT5 data that cannot be identified by disabled pixels in the smaller telescopes.

As FlashCam-type cameras are DC-coupled, the baseline DC level can be measured to give an estimate of the illumination of a pixel, and hence the NSB rate present \citep{FlashcamPerformance}. In principle, this can be read out per event (typically $\mathrm{\sim2.5\,kHz}$), however, due to data storage limitations, this only happens every 0.1\,s for the version of the camera currently installed on CT5. This differs from the older H.E.S.S. cameras, where the NSB rate can be measured with photomultiplier currents. These measurements have historically been used primarily in H.E.S.S. for data quality monitoring, but (for the older cameras) are also explicitly used in \textit{HAP} analyses in determining image cleaning thresholds, and in template-based analyses (further discussed in Section \ref{sec:conc}). Given the optical brightness of satellites, trails can be seen in these data.  Our algorithm for detecting satellite trails comprises the following steps:

    Firstly, the NSB rate files, measured per-pixel in CT5, are processed to retain only pixels exceeding an NSB rate of $\mathrm{900\,MHz}$. This cut was selected to retain trail detections under a wide variety of observational conditions. For a (very) rough conversion estimate, the $m_V=4.9$, $T=20,700\,\mathrm{K}$ star 114 Tauri causes an approximately $\mathrm{7000\,MHz}$ pixel NSB rate in FlashCam at a $\mathrm{50\,^{\circ}}$ zenith angle, but it should be noted that this rate depends on zenith angle, telescope optical throughput and other experimental conditions. Assuming Pogson's ratio holds, this means the limiting magnitude for trail detections in this study is approximately $m_V=8$. 

Secondly, from this dataset, possible satellite candidate pixels are identified by the number ($\mathrm{<100}$) of unique NSB entries for that pixel, per run. This cut excludes bright stars from the dataset, as well as excluding candidate trails where there may be potential instrument issues.

 Thirdly, Individual trails are identified by appending these pixel entries to trail objects if they pass neighbourhood (within an average $\mathrm{50\,cm}$ pixel separation, relative to a $\mathrm{5\,cm}$ edge-to-edge pixel size) and timing co-incidence ($\mathrm{<5\,s}$) checks.

Finally, we perform cuts on these candidate trails to identify likely satellites. The number of unique pixels in the trail must exceed 5 and be below 400, the time difference between the first pixel detection of the trail and the last must exceed $\mathrm{3\,s}$ (excluding meteors). The average of the differences of unique pixel entry timings must exceed $\mathrm{0.3\,s}$ to limit gaps in the trail and prevent spurious clustering. Lastly, the calculated velocity of the trail on the camera must also exceed $\mathrm{0.008\,ms^{-1}}$.\

Our technique can only be applied to data from the FlashCam camera as the intervals between NSB data readouts for the other H.E.S.S. cameras (HESS1/HESS1U/the original CT5 camera) are typically too slow (7/8\,s) compared to the average duration of a satellite trail\footnote{The code used in this analysis, dependent on \textit{HAP} (version 18-11), can be found at \href{https://github.com/Tho-Lang/Satellite-trails}{https://github.com/Tho-Lang/Satellite-trails}}.

\section{Trail detection statistics}
\label{sec:det}

\begin{figure}[t]
    \centering
    \includegraphics[width=\hsize]{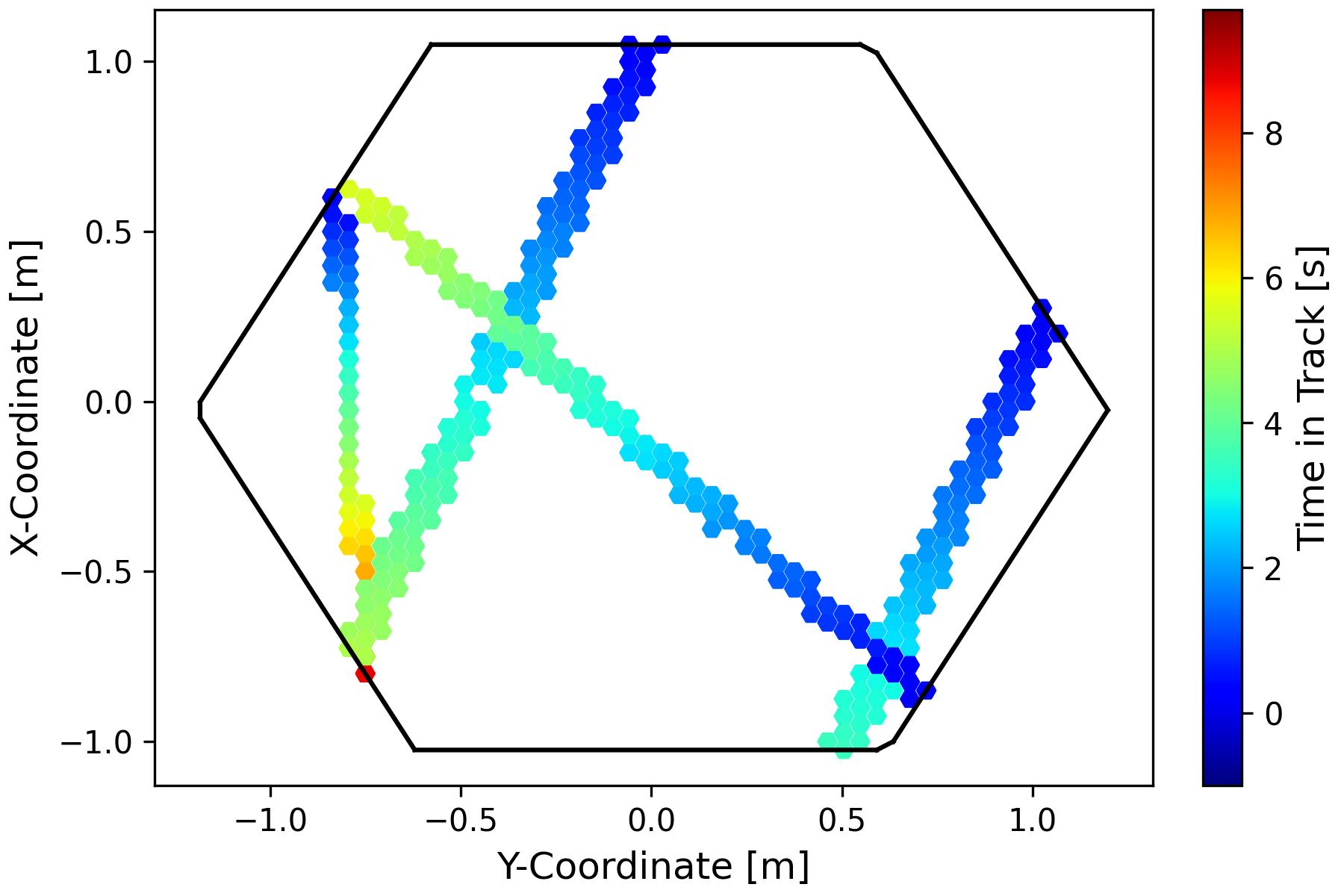}
    \caption{Example of multiple trails in the CT5 camera being detected within a single observing run. The colourbar shows the time since the initial detection of the trail. Often satellites travelling in parallel can be seen on the camera with a short time delay. Only the brightest of these trails created a trail of deactivated pixels in the CT1-4 cameras. In FlashCam, the pixels are configured not to deactivate on the timescales relevant for satellite detection.}
    \label{fig:Camera_multiple_trails}
\end{figure}

The procedure detailed in Section \ref{sec:ident} results in 1658 satellite trail detections in the data. An example of a run with multiple trail detections using our method can be seen in Figure \ref{fig:Camera_multiple_trails}. We observed that parallel trails can occur in such individual runs, consistent with the launch of satellites in `trains' \citep{2020A&A...636A.121Hainaut}. To validate our trail detection technique, a search for satellite trails during the run shown in Figure \ref{fig:Camera_multiple_trails} using the all-sky camera attached to the \textit{ATOM} optical support telescope at the H.E.S.S. site was performed. Three satellite trails were observed in this data, but no bright trails that would unambiguously cross the CT5 FoV during the run were found. This is likely due to the comparatively long interval between all-sky camera images being taken (currently every 3 minutes due to data storage limitations). 

\begin{figure*}
\resizebox{\hsize}{!}
        {\includegraphics{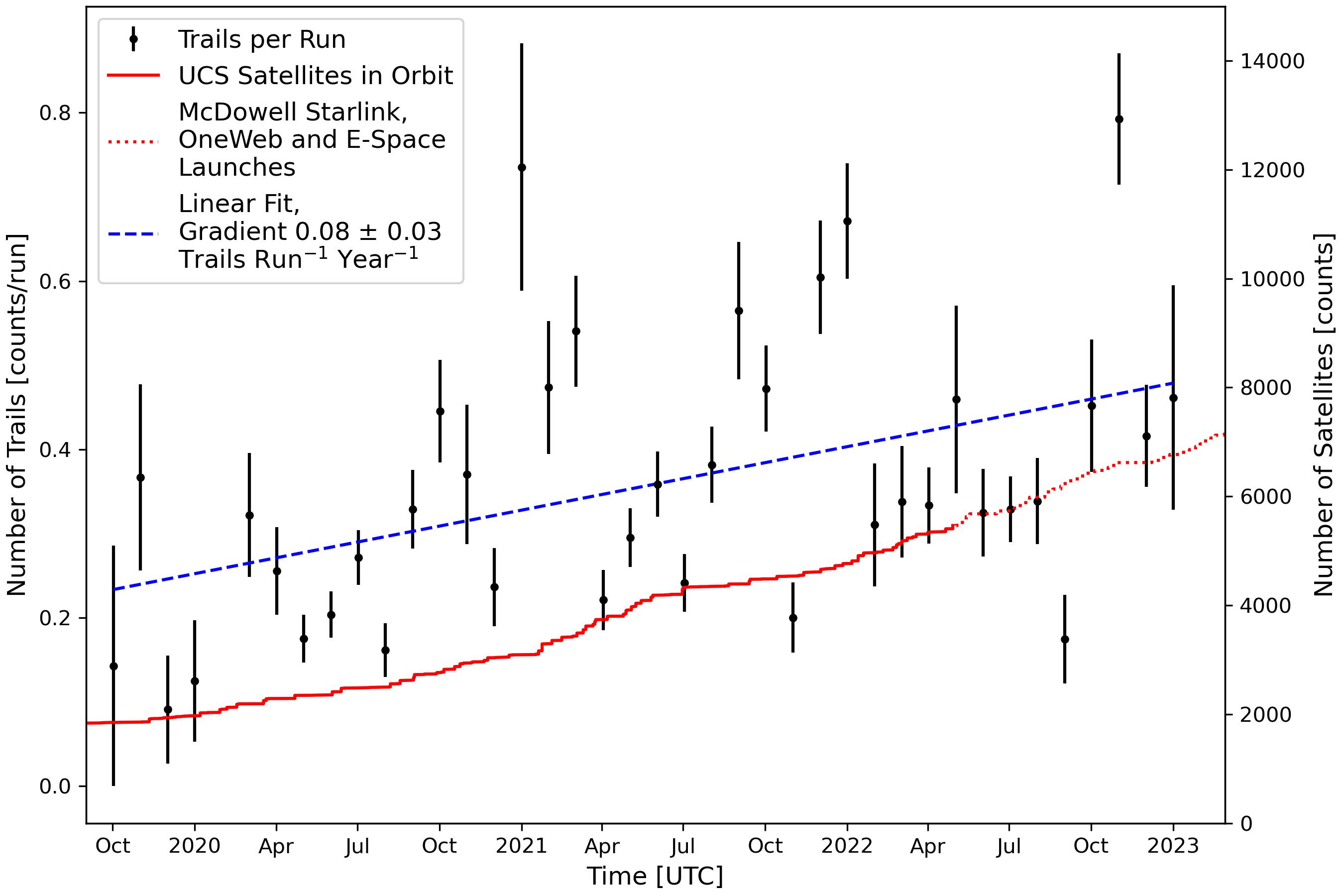}}
    \caption{Number of trail detections (with associated Poisson error) compared to the number of satellites in orbit. The number of satellites in orbit is taken is from \citet{ucs}, but this only extends to April 2022. As such, we augment this data with the number of Starlink, OneWeb and E-Space launches since then (according to \citet{mcdowell}) as an approximate lower limit for the total number in orbit.}
    
         \label{fig:monthly_avgerge}
\end{figure*}

\begin{figure}[h]
    \centering
    \includegraphics[width=\hsize]{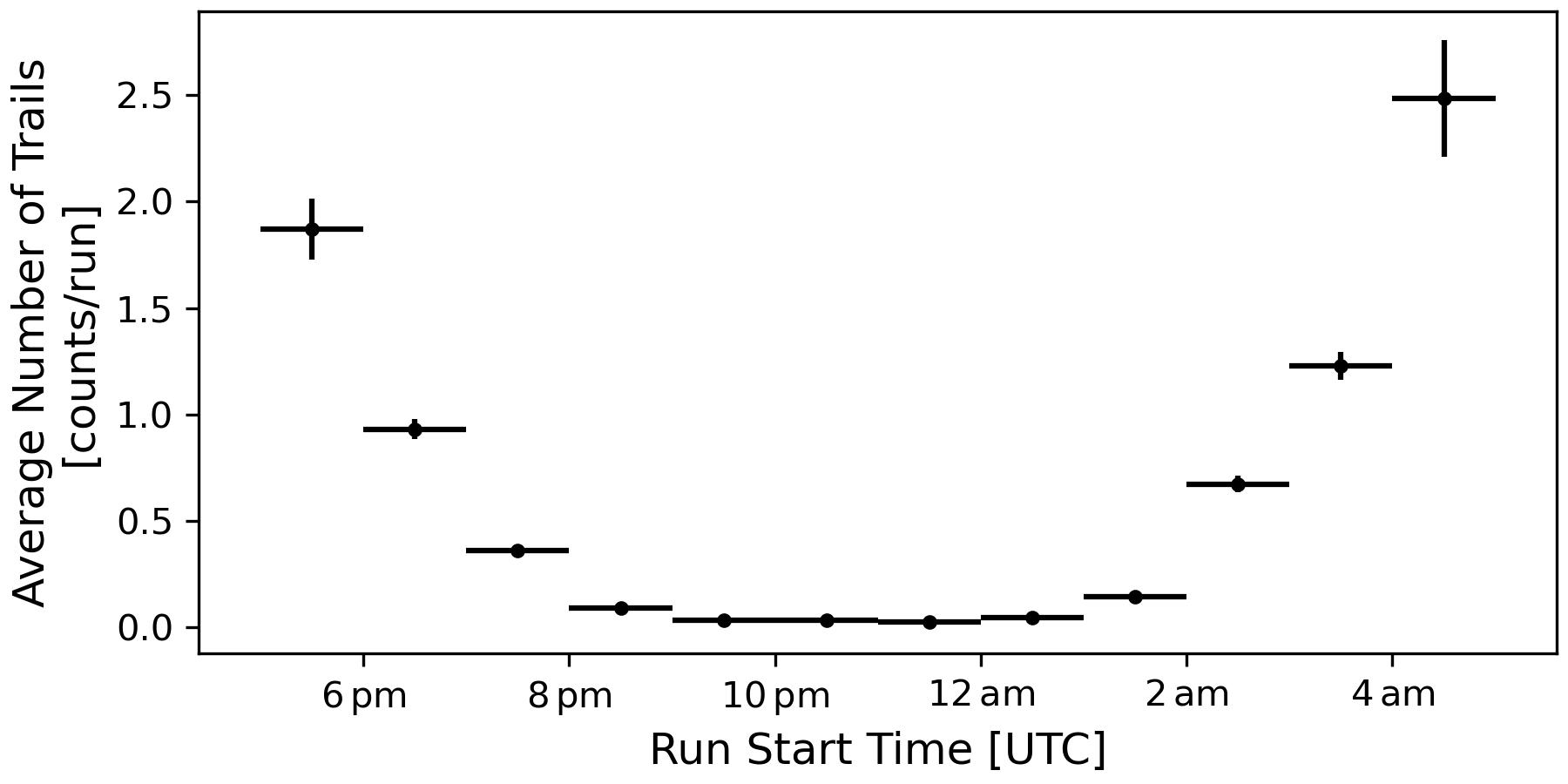}
    \caption{Histogram of the average number of satellite trails we detected as a function of the time at the start of the run (including runs where we detect no trails). Most satellite trails are detected close to the beginning and end of the night when visibility of low-orbit satellites is greatest \citep{2020A&A...636A.121Hainaut,Bassa_2022}. The horizontal errors indicate the size of the bins (also the case for Figures \ref{fig:sunalt_average}, \ref{fig:zen_hist_average}, \ref{fig:velocity_hist_average}, \ref{fig:hist_duration}, and \ref{fig:brightness_hist}).}
    \label{fig:time_in_night_average}
\end{figure}

\begin{figure}[h]
    \centering
    \includegraphics[width=\hsize]{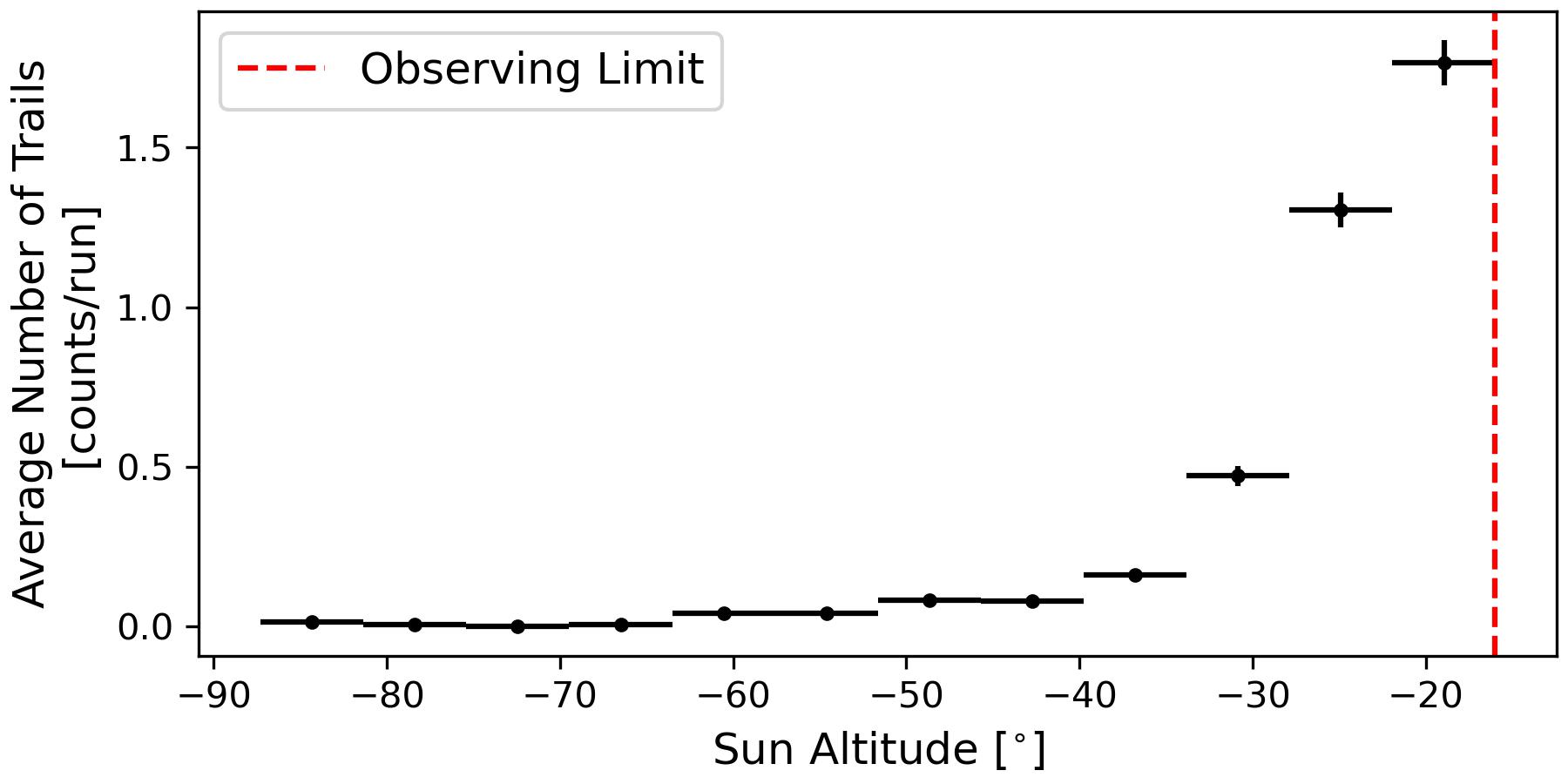}
    \caption{Number of satellites we detect as a function of sun altitude at the beginning of the run. The current sun elevation limit to which H.E.S.S. conducts astronomical observations is $\mathrm{-16\,^{\circ}}$.}
    \label{fig:sunalt_average}
\end{figure}
\begin{figure}
    \centering
    \includegraphics[width = \hsize]{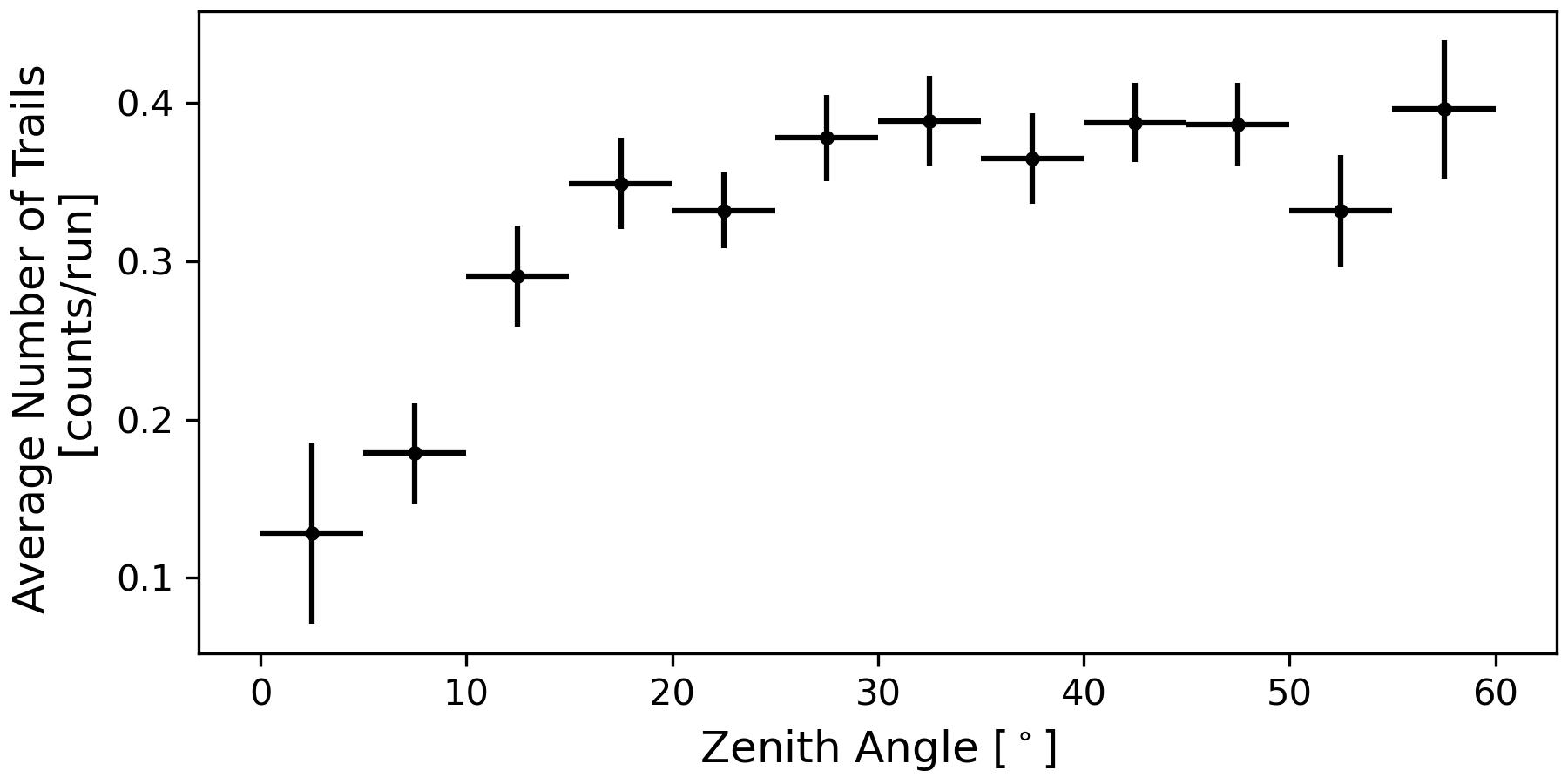}
    \caption{Average number of trails we detected as a function of observation zenith angle, which increases due to the increased size of the sampling volume \citep{2020A&A...636A.121Hainaut,Bassa_2022}.}
    \label{fig:zen_hist_average}
\end{figure}

\begin{figure}
    \centering
    \includegraphics[width = \hsize]{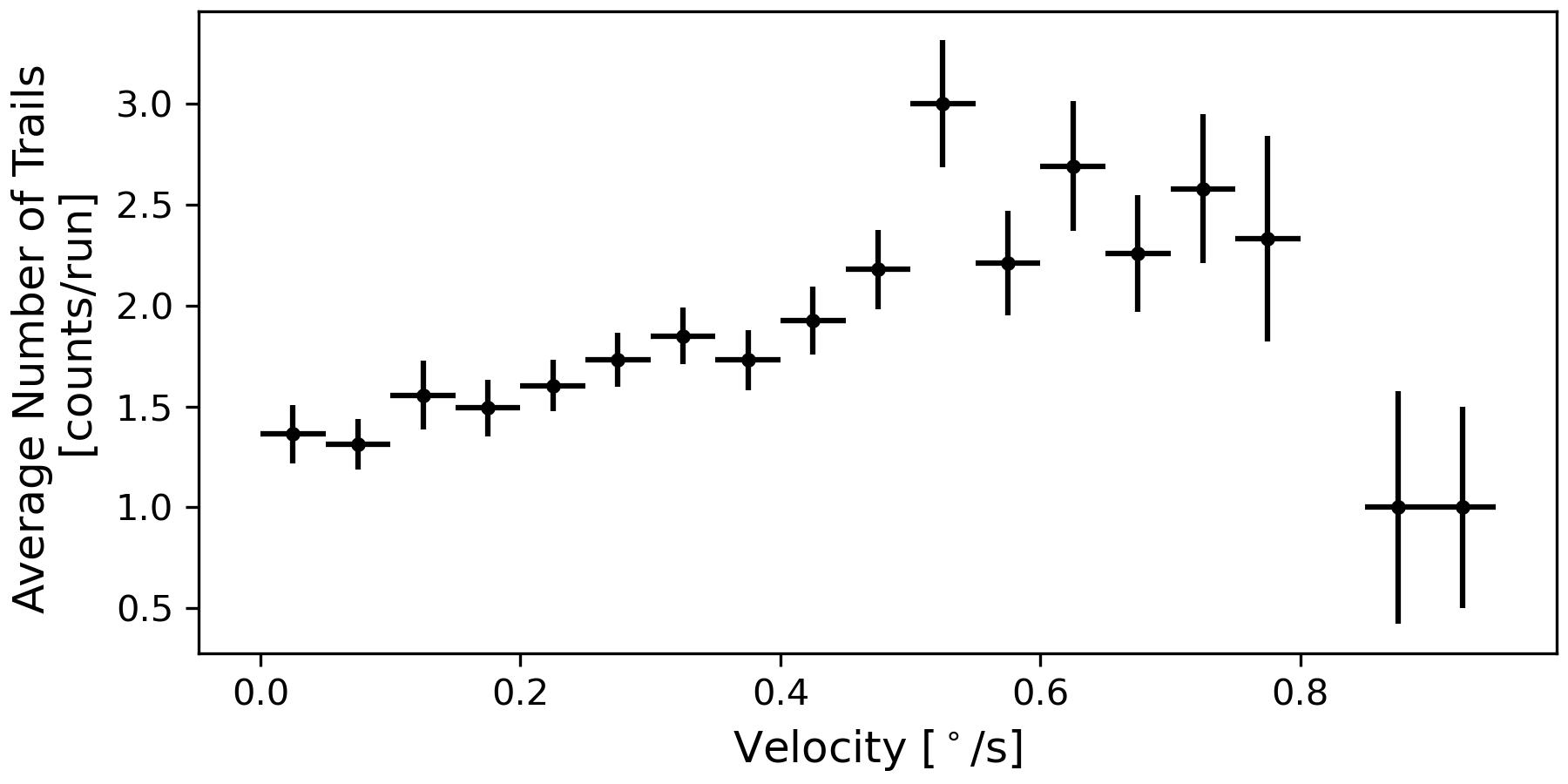}
    \caption{Average number of trails detected as a function of observed trail velocity.}
    \label{fig:velocity_hist_average}
\end{figure}

\begin{figure}[h]
    \centering
    \includegraphics[width=\hsize]{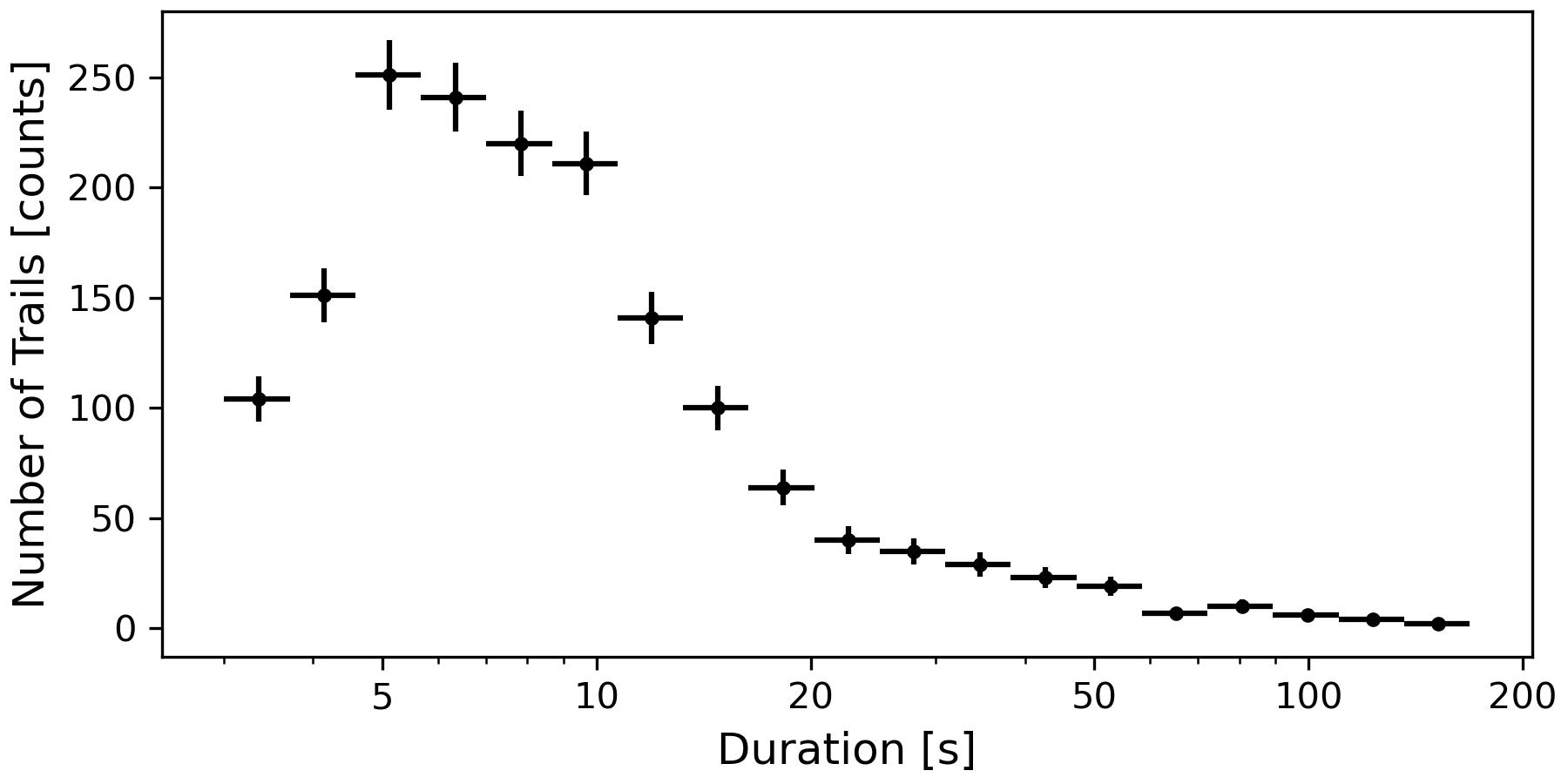}
    \caption{Number of trails detected as a function of trail duration; the observed trail duration does not directly correspond to trail velocity as it is a function of trail position in the camera.}
    \label{fig:hist_duration}
\end{figure}

\begin{figure}[h]
    \centering
    \includegraphics[width=\hsize]{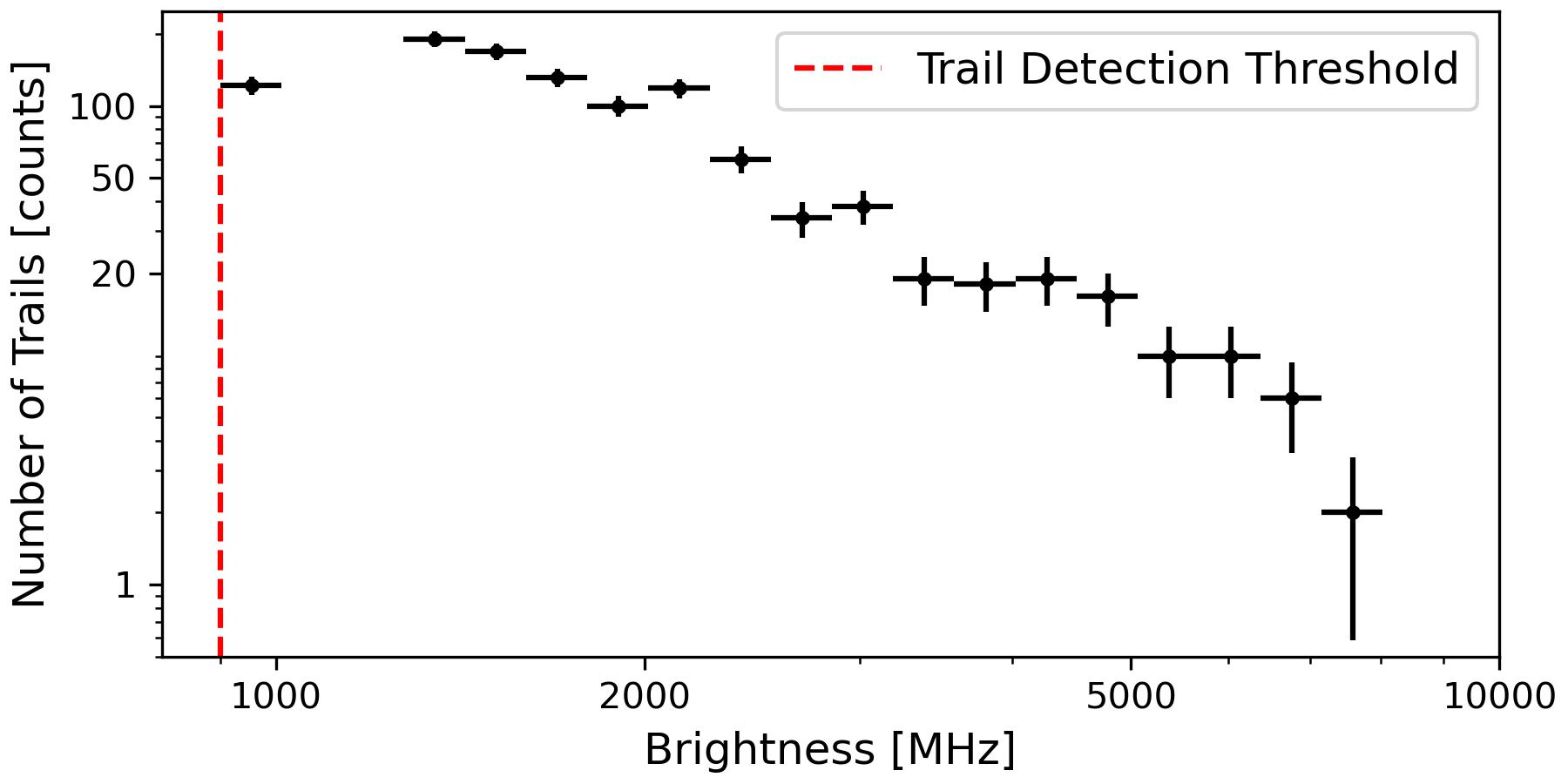}
    \caption{Average trail brightness in MHz, the trail detection threshold of $\mathrm{900\,MHz}$ is shown for comparison.}
    \label{fig:brightness_hist}
\end{figure}

\begin{table*}
\caption{Pearson correlation coefficients between the number of trails detected and observation monitoring data taken during runs.}
\label{table:pearson}      
\centering          
\begin{tabular}{c c}     
\hline\hline
Run Monitoring Variable & Pearson Correlation Coefficient\\ \hline
CT5 Measured Mean Pedestal NSB Rate & 0.086 \\
CT5 Measured Optical Efficiency & -0.073\\ 
Optical Atmospheric Transparency Coefficient & -0.002\\
\hline
\end{tabular}
\tablefoot{Telescope optical efficiency here refers to that obtained by measuring the brightness of muon ring events, for more information see \citet{muons}. The H.E.S.S. atmospheric transparency coefficient variable is described in \citet{Hahn_2014}.}
\end{table*}

Figure \ref{fig:monthly_avgerge} shows the number of satellite trails detected over time. Satellite launch data taken from the \citet{ucs} (UCS) satellite database\footnote{May 1st 2022 version} is also shown for comparison. This data ends in April 2022, so we include launches from the Starlink, OneWeb and E-Space constellations since then as a lower bound for the number currently in orbit \citep{mcdowell}. A linear fit to the data was performed showing a mildly increasing trend, we also only find mild statistically significant correlation between the combined launch dataset and the number of trails we detect (Pearson correlation coefficient 0.41 with a p-value of 0.01). The gradient of this linear fit suggests that the rate of trail detections in H.E.S.S. data has approximately doubled over the time period considered. Satellites in megaconstellations are expected to be brightest in the days immediately following launch prior to them moving to higher orbits \citep{2020A&A...636A.121Hainaut}, but we do not observe a significant spike in satellite trail detections corresponding to the largest launches. This is likely a result of the low likelihood of such `trains' crossing the FoV. Table \ref{table:pearson} shows the Pearson correlation coefficients between the number of trail detections and the H.E.S.S. observational conditions; no correlations are found (as all the values have a magnitude less than 0.1), suggesting our trail detection method (on the selected dataset) is reasonably unbiased with respect to observational conditions.

\begin{figure*}
\resizebox{\hsize}{!}
        {\includegraphics{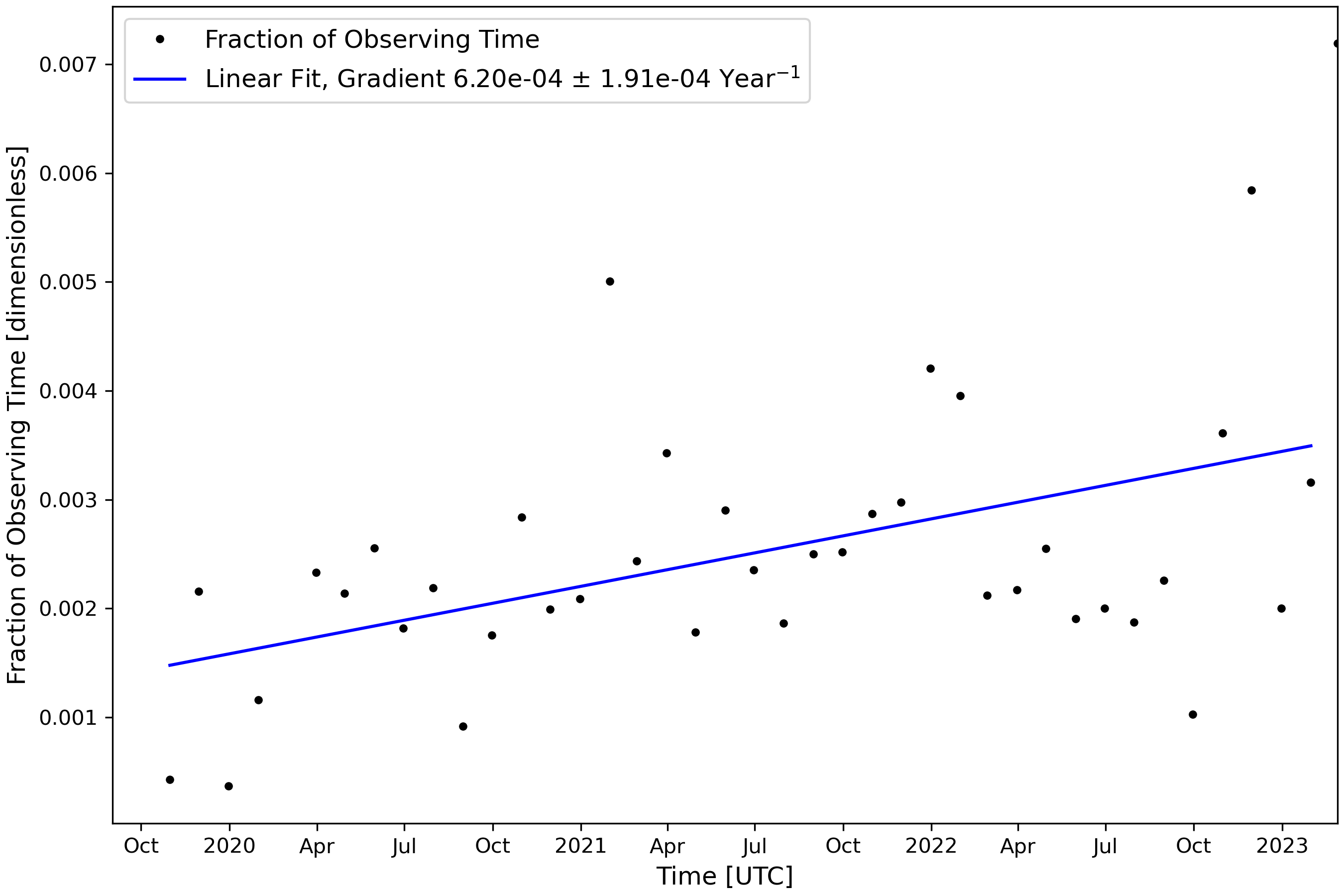}}
    \caption{Fraction of total observing time duration (integrated over all runs which meet the run selection criteria in Section \ref{sec:impact}) affected by satellite trails detectable using our technique.}
         \label{fig:obsfraction}
\end{figure*}

The distributions of trail detections as functions of run start time, sun altitude at the beginning of the run, run zenith angle, trail velocity, trail duration, and trail brightness are shown in Figures \ref{fig:time_in_night_average}, \ref{fig:sunalt_average}, \ref{fig:zen_hist_average}, \ref{fig:velocity_hist_average}, \ref{fig:hist_duration}, and \ref{fig:brightness_hist}. In line with the expectations, we detected significantly more trails near the beginning or end of the observing night (as lower-orbit satellites are more visible), when the sun elevation is higher, and during high-zenith-angle ($\gtrsim25\,^{\circ}$) observations (as the the effective sampling volume increases) \citep{2020A&A...636A.121Hainaut,Bassa_2022}. The plateau in trail rate at an approximately $\mathrm{20\,^{\circ}}$ zenith angle in Figure \ref{fig:zen_hist_average} (despite a steadily increasing number of satellites present) is also expected, as the brightness of Starlink satellites is predicted by \citet{2020A&A...636A.121Hainaut} to drop at this point. There is no direct correspondence between trail velocity and duration; as the trail can intersect the camera at different angles, the distribution of trail durations we observe is more strongly peaked. We observed an approximately power-law distribution in average trail brightness. This also corresponds to expectation as the number of satellites brighter than those from Starlink (such as the recently launched BlueWalker 3, which has been observed to have a peak visual magnitude of 1.4 \citep{mallama2023bluewalker}) is small. Figure \ref{fig:obsfraction} shows the fraction of observational data containing trails over time, which is also slowly increasing and has a median value over our dataset of approximately $0.2\%$. In principle, this fraction of data could likely be removed without a significant effect on H.E.S.S. astrophysical science. Nevertheless, in the next section, we evaluate the impact of the satellite trails on affected events.

\section{Impact on H.E.S.S. gamma-ray data}
\label{sec:impact}
\begin{figure}[h]
    \centering
    \includegraphics[width=\hsize]{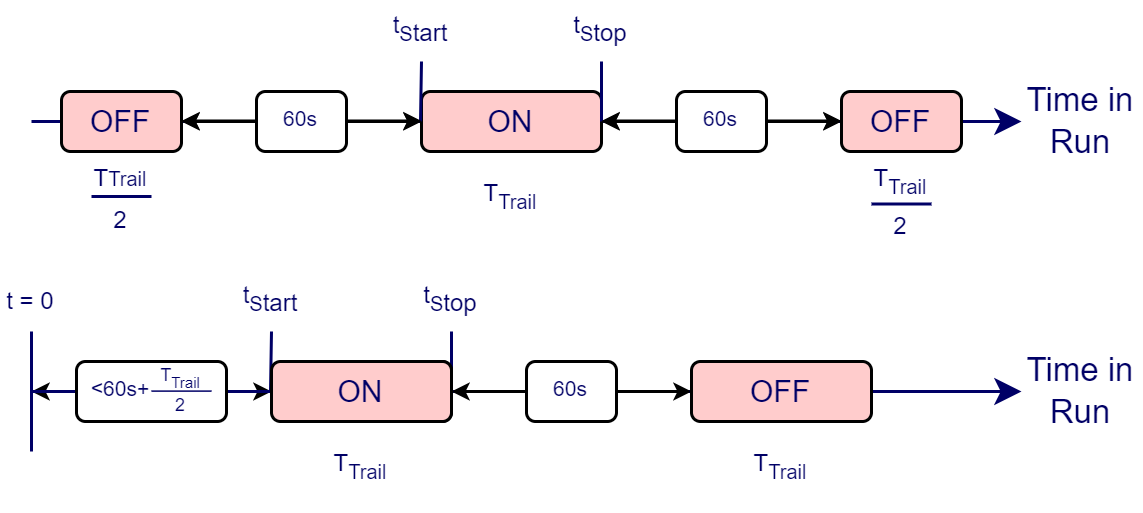}
    \caption{Diagram illustrating the ON and OFF timings relative to the trail start time $\mathrm{t_{Start}}$ and end time $\mathrm{t_{Stop}}$. To account for any possible interference, as well as the larger FoV of CT1-CT4, these timings are separated by 60\,s. If possible, the OFF timing consists of two time periods prior and post trail as shown above, each taking half the duration of the length of the trail $\mathrm{T_{Trail}}$. If not, only a single OFF region is used, either before or after the trail (depending on when the trail occurs in the run). 
    }
    \label{fig:On_off_diagram}
\end{figure}
\begin{figure*}[h]
\resizebox{\hsize}{!}{
    \includegraphics{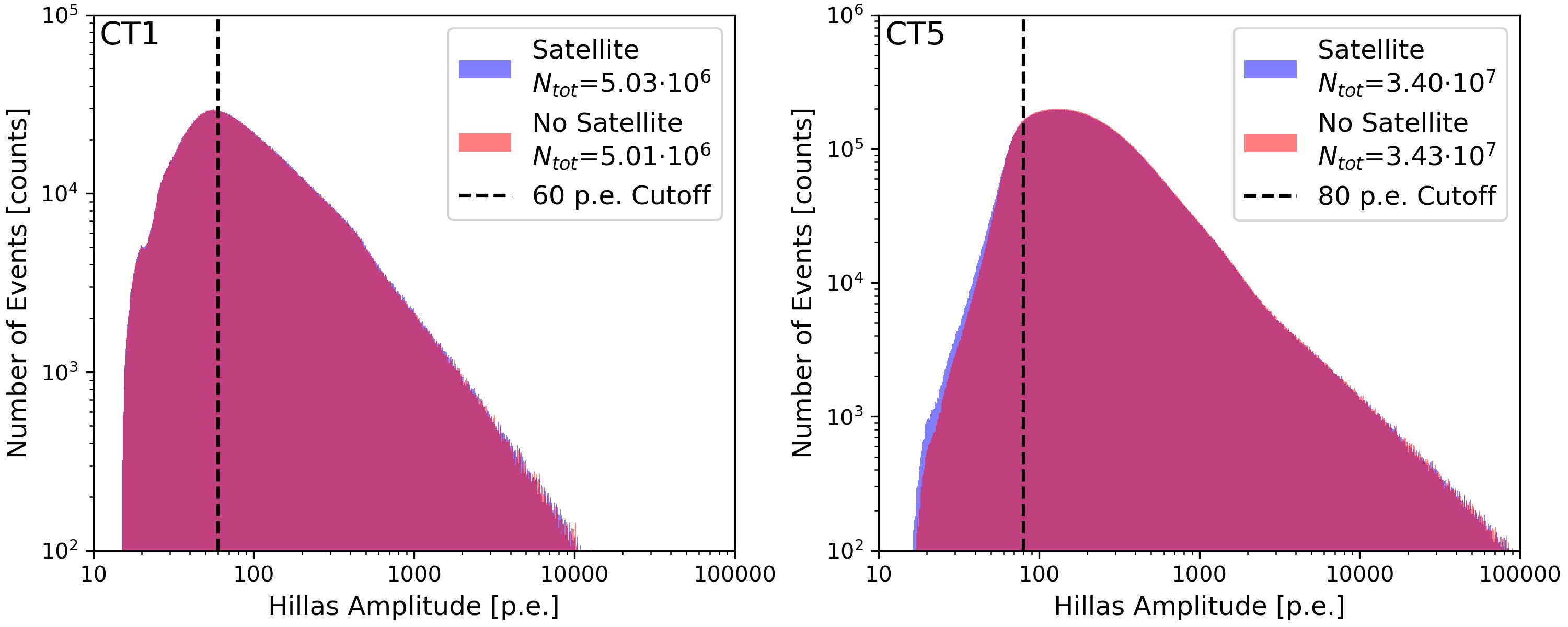}
    }
    \caption{Hillas amplitude distributions for CT1 (exemplary for structurally identical telescopes CT1-CT4) and CT5 for 1557 detected satellite trails with a brightness corresponding to a NSB rate less than 3000\,MHz. For CT5 a small excess of low amplitude events below the hybrid (i.e. CT1-5) standard analysis cutoff of 80\,p.e. can be registered.
    }
    \label{fig:CT1+CT5_amps_dim}
\end{figure*}

\begin{figure*}[h]
\resizebox{\hsize}{!}{
    \includegraphics{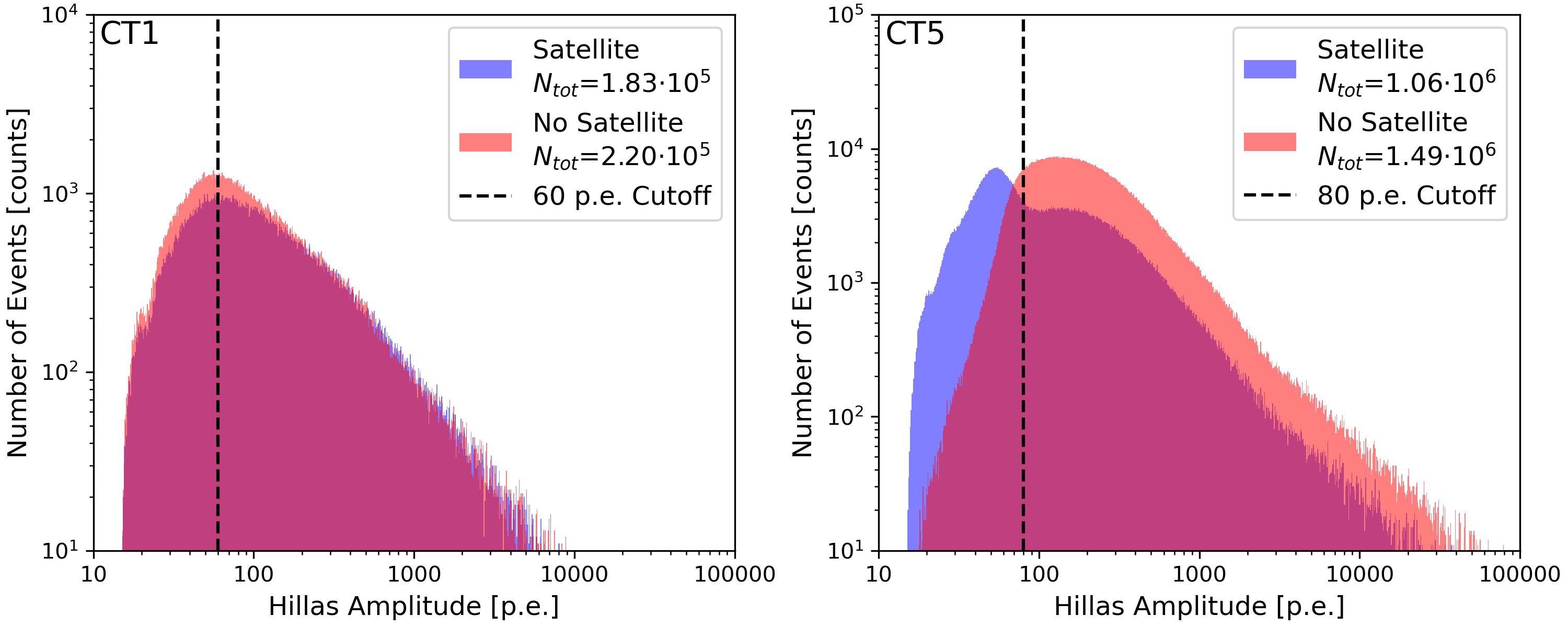}
    }
    \caption{As in Figure \ref{fig:CT1+CT5_amps_dim}, but for 99 trails with a NSB rate higher than 3000\,MHz. A reduction of the event trigger rate can be found in all telescopes, and the excess of low amplitude events in CT5 is now more pronounced.
    }
    \label{fig:CT1+CT5_amps_bright}
\end{figure*}

\begin{figure*}[h]
\resizebox{\hsize}{!}{
    \includegraphics{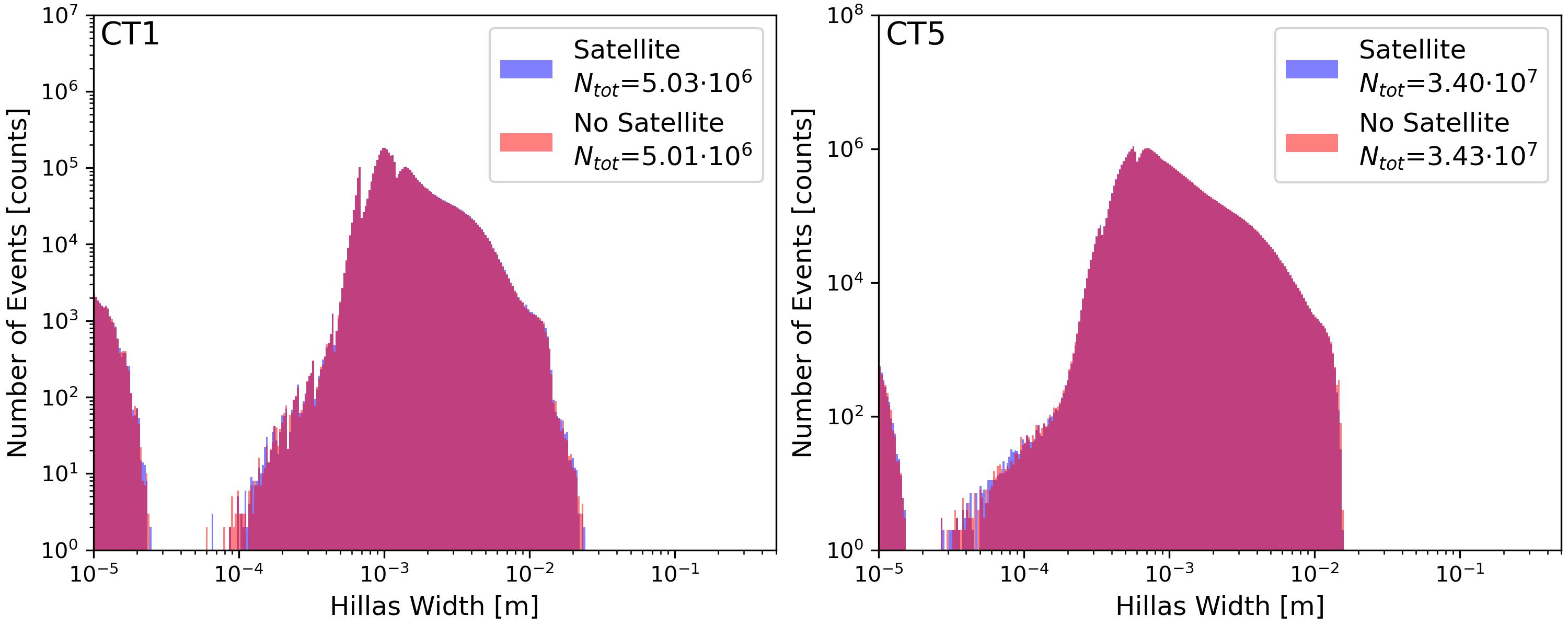}
    }
    \caption{As in Figure \ref{fig:CT1+CT5_amps_dim} for dimmer tracks, except here  Hillas widths are used to quantify the effect on gamma-hadron separation. Only a minimal effect is observed in CT5. Unlike for Hillas amplitude, the configuration used does not impose a cut on Hillas width. Due to the absence of a local distance cut being applied to this data, a population of events truncated by the edge of the camera can be seen with lower Hillas widths.
    }
    \label{fig:CT1+CT5_widhts_dim}
\end{figure*}

\begin{figure*}[h]
\resizebox{\hsize}{!}{
    \includegraphics{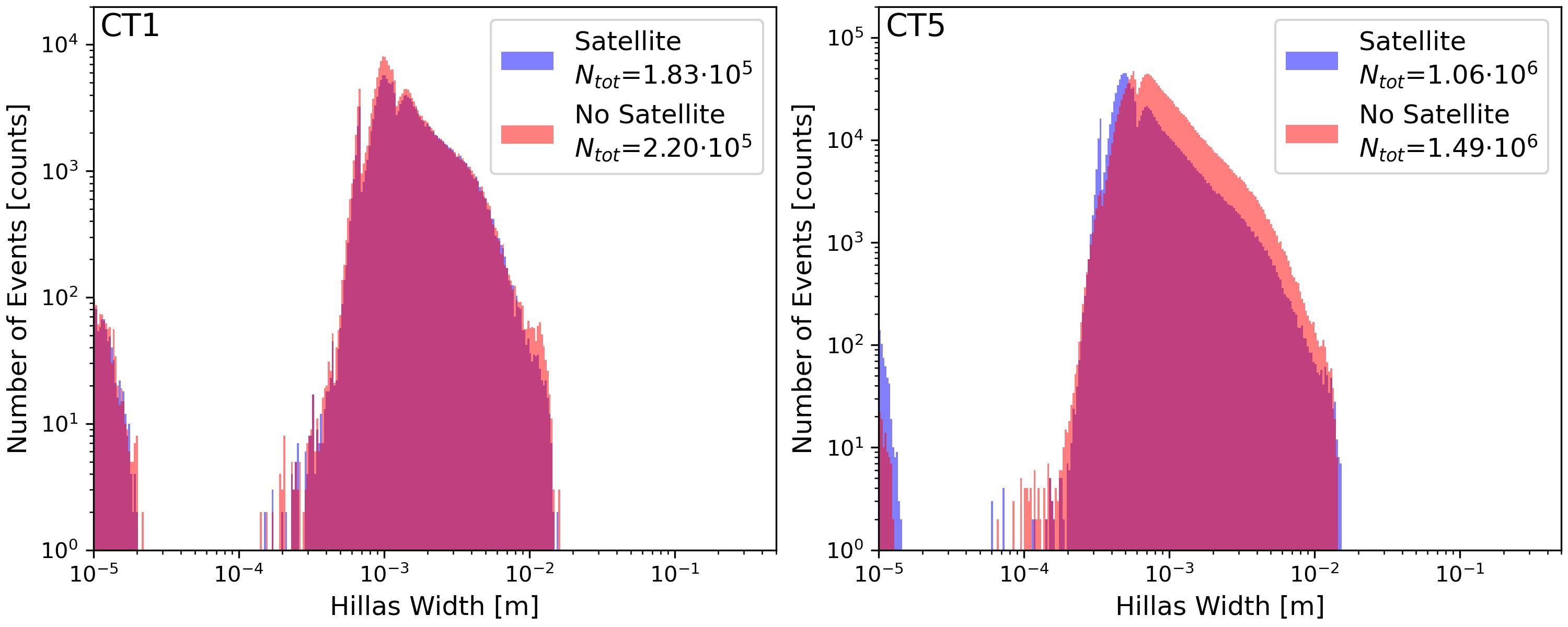}
    }
    \caption{As in Figure \ref{fig:CT1+CT5_amps_bright} for brighter tracks, but here with Hillas widths. A more pronounced effect is observed in CT5, with a decrease in trigger rate. New background rejection methods  aimed at reducing this cut would therefore be affected for CT5-class telescopes, but only in the brightest $\sim5\%$ of trails. The likelihood of observing truncated events in CT5 appears to be increased during satellite trails.
    }
    \label{fig:CT1+CT5_widhts_bright}
\end{figure*}

In this section, we will consider the effect of satellites trails on the event classification and reconstruction used by H.E.S.S.. Depending on observational conditions, we have found that the brightest satellite trails can potentially activate safety mechanisms built into the CT5 camera to prevent the trigger rate from exceeding a safe threshold; this results in a non-linear trail detection response. In these cases, the trigger rate in CT1-4 can increase, but the trigger rate in CT5 can drop significantly. Thus, to appropriately compare Hillas parameter distributions during the trails and at other times, we opted for an approach normalising by time, whereby we defined an ON track and an OFF track time period. This OFF time period is of equal duration to the ON time period but split into two segments, one a minute before the trail starts, and one a minute after the trail ends. This time difference was chosen as it corresponds to the longest duration satellite trails we detected,  hence, selecting this interval ensures the OFF regions will not be contaminated by trail events, but at the same time ensuring that observational conditions are sufficiently similar to make fair comparisons. In cases where the trail was detected within a minute of the beginning or the end of the run, only a single OFF track period was used (illustrated in Figure \ref{fig:On_off_diagram}).

Figures \ref{fig:CT1+CT5_amps_dim}, \ref{fig:CT1+CT5_amps_bright}, \ref{fig:CT1+CT5_widhts_dim}, and \ref{fig:CT1+CT5_widhts_bright} show the Hillas amplitudes and widths observed for the dim trails below 3000\,MHz total brightness and brighter trails respectively for CT1 (as an example of a HESS1U telescope) and CT5. The total duration of the observations we use for ON and OFF data are equal and correspond to 5.37\,hours each. The greatest effects are noticeable in the Hillas amplitudes for CT5, where it appears the brightest satellites can cause `false trigger' events. This is somewhat in contradiction with the results of \citet{2020arXiv200305472Gallozzi} (the only current simulation study that concerns satellites in IACT data), who suggested the need for potentially re-designed trigger logics to avoid false trigger events from satellites, but that the CTA small-size telescopes ($\mathrm{4.3\,m}$ diameter) would be those most affected. Based on the threshold at which such effects become noticeable (3000MHz), this suggests future satellites constellations must keep their brightness fainter than $m_V=6$ in order to avoid affecting event reconstruction for future IACT arrays. The effect on Hillas widths is minimal for both telescope classes, suggesting background rejection against hadronic EAS will not be significantly affected. We performed similar analyses for Hillas parameters that would affect directional reconstruction, however, there was no observable difference for events taken during satellite trails, suggesting that directional reconstruction will not be affected at all by satellites. 

As an example, we compare our Hillas amplitudes to the standard set of high-level Hillas parameter cuts applied at the final analysis stage in \textit{HAP}, the specific values for these are shown in Table \ref{table:cuts}. Given that the false trigger events corresponding to those taken during the brightest satellite trails would fall below this CT5 Hillas amplitude cut, they would not currently be included in most astronomical observations. In cases where these cuts are relaxed to lower the energy threshold of the analysis (so-called loose cuts), as is done, for example,  in observations of gamma-ray bursts with H.E.S.S. (see \citet{Aharonian_2023}), a small percentage of the data (approximately $\sim5\%$ of the observing time fraction shown in Figure \ref{fig:obsfraction}) would likely be contaminated.

\begin{table}
\caption{Typical parameter cuts used during high-level \textit{HAP} analyses.}    
\label{table:cuts}
\centering          
\begin{tabular}{c c c}     
\hline\hline

Hillas Parameter & Cut Values (Lower, Upper) & Unit\\ \hline
CT1-4 Amplitude & (60, 1000000) & p.e. \\
CT5 Amplitude & (80, 1000000) & p.e.\\   
CT1-4 Local Distance & (0, 0.525) & m \\ 
CT5 Local Distance & (0, 0.72) & m \\
\hline
\end{tabular}
\end{table}

\section{Future predictions and strategies to mitigate impacts}
\label{sec:pred}
Since no CTA telescope will consist of an identical instrument configuration to any of the H.E.S.S. telescopes, it is difficult to draw conclusions from our study that would be specifically applicable to CTA. We can, however, make approximations based upon the FoV of the IACT telescopes, assuming the number of trails scales quadratically with the FoV (the number of satellites that can be identified as trails in the NSB data may scale linearly, but the number that can influence the near-instantaneous Cherenkov data will scale quadratically). By extrapolating the linear model shown in Figure \ref{fig:monthly_avgerge} (based on our 1658 trail detections), we can make a (rough) prediction for the number of satellite trails that would be present in the data per-hour, and the percentage of observing time affected for the three classes of CTA telescopes at the beginning of 2030. We assume that the satellite orbit and altitude distributions remain the same as today. These predictions are shown in Table \ref{table:pred}. Given the three values in this table, and assuming no further mitigation strategies are used, it appears likely that the fraction of observation time affected by satellites for the next generation of IACTs will be comparable to the amount of duty cycle lost due to instrument issues in the current generation ($\sim2.5\%$). But given the results in Figure \ref{fig:CT1+CT5_amps_bright}, only a $\sim5\%$ fraction of the trail co-incident data would be sufficiently affected to justify its removal from scientific analyses. The trends we observe with H.E.S.S. may differ for IACT arrays with alternative instrumental designs, observation strategies, or locations; thus, we recommend other current generation IACT collaborations perform similar studies to ours to independently validate these results. The data we have collected should also be useful for developing observing strategies for other astronomical instruments present or planned in the Khomas Highland, including HATSouth \citep{HATSouth} and the African Millimeter Telescope \citep{AMT}.

\begin{table*}
\caption{Rough predictions for the number of trails seen per-observing-hour and the fraction of observational data per-telescope affected in 2030 for the three CTA telescope classes.}
\label{table:pred}      
\centering          
\begin{tabular}{c c c c}     
\hline\hline
Telescope Class & FoV Size [$^{\circ}$] & Average Trails Per Hour [counts] & Observing Time  Affected [\%]\\ \hline
Small & 9.0 & 15 & 5\\
Medium & 8.0 & 12 & 4\\   
Large & 4.3 & 3 & 2\\
\hline
\end{tabular}
\tablefoot{These are based on the assumption that the linear trends (fitted to the 1658 trails we detect) shown in Figures \ref{fig:monthly_avgerge} and \ref{fig:obsfraction} continue, that no mitigation measures are taken, and scaling relative to the FoV of the current CT5 camera ($3.4\,^{\circ}$). These estimates are subject to large uncertainties with respect to future satellite launches, instrument design and also significant extrapolation from current data, and as such should be treated with caution.}
\end{table*}

Recently, IACT arrays have commenced observations using intensity interferometry, which utilises the second order coherence of light to perform high resolution optical astrometry \citep{Abeysekara_2020_SII,Acciari_2019_SII}. Temporal correlations between the arrival times of optical photons at multiple telescopes can be used to sample the u-v interferometry plane, from which image reconstruction can be performed \citep{Dravins_2013}. Intensity interferometric observations are conducted using a single pixel per telescope, such that the FoV is much reduced, however continuous measurements are required over many minutes or even hours. 
As such, the likelihood of a satellite traversing the pixel FoV is much lower, however should this occur, the data collection will be much more strongly affected. The most efficient mitigation strategy in this case would be to simply exclude all data from times at which a satellite traversed the pixel FoV. Fortunately, this is feasible offline, as the temporal correlation is not, in any case, usually performed during data acquisition \citep{DRAVINS2013331}.  

Observations with IACTs are a mixture of pre-scheduled runs and followup of transient objects, that is, so-called target of opportunity (ToO) observations. Planned observations of persistent sources could (in principle) be scheduled to avoid such contamination by satellite trails, but this is not the case for ToOs. We do not observe a systematic bias in the number of trails detected during ToO observations; the ratio of trails we detected during ToO runs compared to the overall number of trails is the same as the ratio of runs targeting ToOs out of all observing runs ($\sim16\%$). Software to predict the presence of satellite trails during astronomical observations is being developed \citep{CUI2022100652}, although attempting to calculate the position of every LEO object to avoid satellite trails completely during observations is likely to be computationally prohibitive \citep{Bassa_2022}. Other mitigation options would be simpler strategies to simply not observe west in the first hour of the night and east in the last \citep{Bassa_2022} or to include weighting factors in IACT schedulers. These weighting factors would balance the likelihood of a transit occurring in the IACT data (and some of the data needing to be excluded) against the potential science gain of that particular IACT observation, similar to an approach investigated for the Vera Rubin Observatory \citep{Hu_2022}. However, it appears the (low) fraction of H.E.S.S. observing time affected by such trails would not justify either of these strategies for H.E.S.S. or CTA.

\section{Discussion and conclusions}
\label{sec:conc}

In \textit{HAP}, an additional step of event energy and direction reconstruction, called an image pixel-wise fit for Atmospheric Cherenkov Telescopes (ImPACT), is applied following Hillas-parameter-based analysis. This is based on likelihood fitting to templates based on Monte Carlo simulations in order to improve performance \citep{Parsons2014}. This fit is seeded with the reconstructed direction and energy values from multiple Hillas-parameter-based event reconstructions with differing cleaning thresholds. In theory, false Hillas amplitude values due to satellite trails could therefore propagate through to affect such template-based analyses. New deep-learning-based event classification and reconstruction methods \citep{shilon}, or analyses that treat different quality events in separate classes \citep{Hassan_2021}, could also be affected by satellite trails more severely. This is as they aim to reduce the need for Hillas parameter cuts, either for the purpose of reducing the energy threshold of the analysis, or to process events truncated by the edge of the camera. It should be noted that we also do not explicitly consider the role of satellite detections in IACT pointing calibration, for which images of stars are used. False peaks in such images (taken with either CCDs or IACT cameras themselves with dedicated slow-signal chains) could cause pointing reconstruction errors on the longer integration timescales of seconds used for such observations. This should be investigated in a further study, though corrections for this could be performed offline. Also not considered here, satellite trail identification could be performed in online analysis to remove satellite-contaminated data in near-real-time, to reduce the necessary data transfer for CTA. Finally, given that they effectively increase the instrument FoV for the purposes of transient follow-up, CTA observations in divergent pointing mode could be more affected more severely than traditional analyses \citep{CTADivergent}, which we have not addressed in this study.

The potential scientific loss associated with the launch of satellite constellations must be balanced against their potential societal benefit. The ability to provide reliable internet connections to remote regions (where providing conventional internet infrastructure is economically unfeasible) has a number of potential advantages, including for socio-economic development, education and healthcare \citep{starlinkafrica}. Notably, in Namibia where H.E.S.S. is situated, only 41 \% of the population use the Internet \citep{wb:2021}. Even for astronomers, starlink-type internet provision could be advantageous, as observatories are often themselves located in remote regions. H.E.S.S. itself could benefit in this regard; a Starlink-based internet connection might provide improved stability for data transfer from the telescope site to data processing facilities in Europe at a lower cost. The potential benefits of satellite-constellation-based internet provision must also be weighed against the cultural and economic significance of dark skies in developing countries, as recently studied with respect to Namibia \citep{2021ASPC..531...17Dalgleish}.

In the present study, we have developed and validated a new method for determining the presence of satellite trails for IACTs and utilised it to quantify the impact of satellite constellations on H.E.S.S. data. Whilst they are measurable, we confirm the ultimate effect of satellite trails on H.E.S.S. observations and data is minimal (see Figures \ref{fig:CT1+CT5_amps_dim}, \ref{fig:CT1+CT5_amps_bright}, \ref{fig:CT1+CT5_widhts_dim}, and \ref{fig:CT1+CT5_widhts_bright}). The fraction of observing time affected is low (Figure \ref{fig:obsfraction}) and standard H.E.S.S. Hillas parameter cuts would likely remove the worst affected events.  With our instrument configuration, observing location, and scheduling strategy, we only observed a slow increase in the number of satellite trails detected over time (Figure \ref{fig:monthly_avgerge}), and a mild correlation between the number of trails we detect and the number of satellite launches. That said, our results show that IACTs with a larger diameter are affected to a greater extent, thus suggesting that new analysis techniques designed specifically to lower the energy threshold of large IACTs could be still be affected. Despite the results of our study indicating it will, in fact, prove to be a minor issue, given the cost of developing CTA and the rapid evolution of this technology, the effect of satellite constellations should continue to be studied and monitored.

\begin{acknowledgements}
This work has been through review by the H.E.S.S. collaboration, who we thank for allowing us to use low level H.E.S.S. data in this work, and for useful discussions with collaboration members regarding this paper. We thank Felix Jankowsky for providing data from the H.E.S.S. all-sky camera monitoring system. This work is supported by the Deutsche Forschungsgemeinschaft (DFG, German Research Foundation) – Project Number 452934793.
\end{acknowledgements}

%
\bibliographystyle{aa} 
\bibliography{references} 
%

\appendix
\onecolumn
\section{Data Release}
Data describing the satellite trails detected and associated observational conditions is provided using CDS, as detailed in Table \ref{table:cds}.

\begin{table}[!htbp]
\caption{Description of data released on CDS.}
\label{table:cds}      
\centering
\begin{tabular}{c c c}     
\hline\hline
Column Name & Unit & Description\\ \hline
Run ID & - & H.E.S.S. run number\\
Track ID & - & Track identifier within run\\  
Mean tail brightness & MHz & Mean of NSB map entries attached to the trail\\
Trail duration & s & -\\
Maximum trail brightness & MHz & Maximum of NSB map entries attached to the trail\\
Trail start time& s & First trail detection timestamp\\
Number of pixels affected & counts & Number of unique pixels attached to the trail object\\
Velocity of satellite & $\mathrm{^{\circ}/s}$ & Measured velocity for the trail\\
Mean run zenith angle & $\mathrm{^{\circ}}$ & -\\
Run duration & $\mathrm{s}$ & - \\

Run start time & UTC & -\\
Mean atmospheric transparency coefficient&-&As defined in \citet{Hahn_2014}\\
Mean run RA & $\mathrm{^{\circ}}$ & Mean run right ascension\\
Mean run Dec & $\mathrm{^{\circ}}$& Mean run declination\\ 
Mean run NSB & MHz & Mean NSB value measured for the CT5 camera during the run\\
Relative mean CT5 efficiency & - & \makecell{Mean CT5 muon efficiency during run \citep{muons},\\ relative to the maximum value in the dataset}\\
\hline
\end{tabular}
\end{table}
\end{document}